\documentclass{optica-article}

\journal{opticajournal} % for journals or Optica Open

\articletype{Research Article}

\usepackage{lineno}
\usepackage{xcolor}

\linenumbers % Turn off line numbering for Optica Open preprint submissions.

\begin{document}

\nolinenumbers

\title{Spectral signatures of coexisting isolas and four-wave mixing under the effect of noise
in photonic oscillators}

\author{Johanne Hizanidis,\authormark{1,*} Vassilios Kovanis,\authormark{2}}
\address{\authormark{1}Department of Physics, University of Crete \& Institute of Electronic Structure and Laser, Foundation for Research and Technology-Hellas, 70013 Herakleio, Greece\\
\authormark{2}Virginia Polytechnic Institute and State University, 
National Security Institute, 900 North Glebe Road, Arlington, Virginia 22203}

\email{\authormark{*}hizanidis@physics.uoc.gr} %% email address is required; see note below about the corresponding author designation

% use {asbstract*} to suppress the copyright line. Copyright information will be added in production

\begin{abstract*} 

The optical power spectrum is the prime observable to dissect, understand, and design the long-
time behavior of small and large arrays of optically coupled semiconductor lasers. A long-standing
issue has been identified within the literature of injection locking in photonic oscillators: first how
the thickness of linewidth and the lineshape spectral envelope correlates with the deterministic
evolution of the monochromatic injected laser oscillator and second how the presence of noise and
the typically dense proximity in phase space of coexisting limit cycles of the coupled system are
shaping and influencing the overall spectral behavior. In addition, we are critically interested in the
regions where the basin of attraction has a fractal-like structure, still, the long-time orbits are P1 (period 1)
and/or P3 (period 3) limit cycles. Numerically computed evidence shows that, when the coupled system lives
in the regions of coexisting isolas and four-wave mixing (FWM) limit cycles, the overall optical power spectrum is deeply imprinted
by a strong influence from the underlying noise sources. A particularly intriguing observation in this
region of parameter space that we examine is that the isolas draw most of the trajectories
on its phase space path.

% , in our case, period-one (P1) and period-three (P3). 
% Via numerically integrating the dynamical system, 

% Additionally, a qualitative comparison is made between the stochastic Langevin simulations and the recorded optical power spectra taken from legacy experiments that we revisit. Finally, we point to a possible set of next-generation applications in photonics and quantum technologies where such findings may have implications.

\end{abstract*}

%%%%%%%%%%%%%%%%%%%%%%%%%%  body  %%%%%%%%%%%%%%%%%%%%%%%%%%
\section{Introduction}
Injection-locked semiconductor oscillators have emerged in the past three decades, as engines for dissecting experimentally and theoretically intricate nonlinear dynamics phenomena, including routes into chaos, generating quasiperiodic signals, and finding the isochrons buried in the periodic limit cycles~\cite{ERN10,OHT13,WIE05,Costas2020,Herrera2021,Jason2010,SPIE2023,simpson1995,Yamamoto4,Varangis1996,WAN15}.  
They have also furnished multiple flexible photonic devices (tabletop and chip scale) to generate low-noise ultra-fast microwave signals and establish a versatile environment for highly tunable photonic oscillators, develop resources for observing quantum mechanical squeezed states used in applications of quantum gyroscopes,  sensitive measurements of gravitational wave detection%\cite{LIGO}% 
and building maps of the Ising model into injection-locked geometry for use in photonic quantum processors~\cite{Yamamoto1,Yamamoto2,Yamamoto3,Yamamoto4,Yamamoto2011}.

In this manuscript, we squarely focus on a perennial topic, namely the interplay of periodic coexisting limit cycles and the underlying noise, to find out the long-time spectral features and quantify the spectral purity of such regions of the injection optical detuning regions. Inspired by older works on high-frequency driven microwave circuits~\cite{DYK90} and micromechanical oscillators~\cite{STA06}, where super-narrow spectral peaks were computed as an outcome of coexisting periodic limit cycles, our investigation, in contrast, is executed on an ultra-fast photonic oscillator that can oscillate at even larger speeds of $80$ GHz.

An attractor of a dynamical system is a subset of the state space where the orbits originating from typical initial conditions tend as time evolves. It is a rather common feature for nonlinear dynamical systems to have multiple attractors. For each such dynamical state, its basin of attraction is the set of initial conditions that eventually lead to it. As a result, the qualitative behavior of the long-time evolution of a given system can be radically different depending on which basin of attraction the initial condition resides in. To the best of our knowledge, there have been very few experimental in-depth studies of basins of attraction because following the evolution of initial conditions in low-frequency macroscopic systems is usually very time-consuming and system parameters tend to drift over the course of many data recording runs~\cite{Roukes2007}. Fractal basin boundaries~\cite{MCD85}, in particular, have been found numerically and investigated experimentally in man-made systems where the key oscillating frequency is rather slow.

A system that is of significant technological relevance and is well-known for demonstrating multiple attractor behavior is the optically injected laser. The complexity of this system involves a rich set of qualitatively different dynamical features, including stable and unstable steady states (fixed points) and self-sustained oscillations (limit cycles), as well
as self-modulated quasiperiodic (tori) and chaotic orbits (strange attractors)~\cite{ERN10,OHT13,WIE05}.
The long-term dynamics of the injection-locking architecture enabled with semiconductor lasers, such as quantum well, quantum dot, and quantum cascade lasers on photonic integrated circuits or on a tabletop configuration with discrete devices have been investigated analytically, numerically, and experimentally for the past 50 years~\cite{Costas2020}. The speed of the free-running relaxation oscillation of these tunable photonic oscillators typically starts at a few GHz and may cross 84 GHz under strong injection and for large optical frequency detuning~\cite{Herrera2021}. Therefore the mapping of the basins of attraction in such a system of ultra-fast oscillators can be explored only numerically. Examples of such explorations and fine detailed numerical scans on regions of chaos and period 3 limit cycle orbits laying within the chaotic orbits, were reported for zero optical detuning in~\cite{Jason2010}.

Semiconductor lasers are known to have a high level of intrinsic noise primarily
due to spontaneous emission~\cite{AGR90,WAN18,DIN23} which broadens and obscures their 
spectral properties. From a dynamic point of view, such fluctuations may lead to noise-induced 
attractor switching and therefore affect the output waveform. Self-sustained oscillators that support robust limit cycles have a major technological significance with devices that generate periodic signals at an inherent frequency and are often engineered to enable highly accurate time or frequency references. Therefore it is of high importance to map out the dynamical regimes of such states and study how stochastic fluctuations interact with their basins of attraction. The prime observable through which these phenomena can be analyzed is the optical power spectrum. One of the key technical issues that we are working forward to is to address the spectrum congestion and demand for higher data rates that are driving a push toward higher carrier frequencies in wireless communications, necessitating sources of exceptionally pure widely tunable radiofrequency carrier signals. Multiple applications may be influenced such as laser radar applications, optical metrology, and spectroscopy. It is worth noting that pivotal issues of current applications of quantum sensing are deeply dependent on the use of properly sharp linewidth as well as wide frequency tunable oscillators~\cite{QuantumGYRO2023} and a range of geolocation applications~\cite{SPIE2023}.

This paper is organized as follows: after the introductory section, the single-mode optical injection rate equation system is presented with noise terms incorporated. Next, we perform a detailed analysis of the deterministic dynamics in the regime of interest where the system's behavior is dominated by the coexistence of four-wave mixing and period-three limit cycles. The corresponding basins of attraction are computed and their fractal-like structure is noted. The following section focuses on the optical power spectra corresponding to the observed dynamics, while the interplay of noise and the fractal-like basins of attraction and its imprint on the spectral properties of the emitted signal is discussed. 
In the conclusions section, we summarize our main findings and 
propose new directions for further studies.

\section{Injected Photonic Oscillator Model}
\label{sec:model}
The dynamic time evolution of an optically injected semiconductor laser subject to an externally imposed  monochromatic signal is modeled, in dimensionless form, by three single-mode rate equations, one covering the dynamics of the amplitude $R$ of the electric field emitted out of the slave cavity, one for the phase offset $\phi$ between the two lasers, and one for the electronic carrier density into the nonlinear gain medium of the slave laser cavity, $N$:
\begin{eqnarray}
\label{eq:rate_eq_R}
\frac{dR}{dt} &=& NR+\eta\cos\phi +F_{a}\\ 
\label{eq:rate_eq_phi}
\frac{d\phi}{dt} &=& \Omega-\alpha N-\frac{\eta}{R}\sin\phi+\frac{F_{p}}{R}\\
\label{eq:rate_eq_N}
T\frac{dN}{dt} &=& P-N-P(1+2N)R^2.
\end{eqnarray}
In the above system, $\eta$ is proportional to the amplitude of the externally
injected field, $\Omega$ is the optical frequency detuning between the slave and master emission optical frequencies, $\alpha$ is the linewidth enhancement factor, $T$ is the ratio of the carrier to the photon lifetimes, $P$ denotes the electronic pumping current above the solitary laser threshold, and $t$ is the time normalized to the photon lifetime. In addition, fluctuations for the amplitude and the phase that the optical gain medium is generating into the slave cavity are incorporated.  These fluctuations are represented by the Langevin noise sources $F_a(t)$ and $F_{p}(t)$, respectively. They are zero-mean, $\langle F_i(t)\rangle=0$, delta-correlated, $\langle F_i( t) F_j(t') \rangle=D\delta_{ij}\delta(t-t')$ ($(i,j=a,p)$) and $D$ is the noise intensity, proportional to the spontaneous emission rate~\cite{PhysicsLetters1998}. 

This set of three nonlinear equations collects and fuses intelligently all the key features of the optical gain of the slave laser cavity and the stochastic dynamics of the optically injected laser, bringing into focus the two-time scales of the electrons (nanoseconds) and the photons (picoseconds), the strong phase-to-amplitude coupling via the linewidth enhancement factor, and the noise that drives and promotes the free-running relaxation oscillation sidebands of the slave laser~\cite{Vahala1983}.

This radically simple-looking rate equation model is well-established and has been the subject of investigations for more than five decades. It has been used to examine the bifurcation structure~\cite{WIE05}, design multiple laser experiments on the tabletop and chip scale configurations, and spin out a whole host of microwave photonic applications including wide-frequency tunable photonic oscillators~\cite{Herrera2021}, bandwidth-enhancing emitters~\cite{simpson1995}, and low-noise microwave oscillators~\cite{Simpson2014}. Also, it has provided a framework for generating novel quantum states such as squeezed quantum mechanical states for injecting quantum light into interferometers, and laser ring gyros, and establishing next-generation phase modulators for hacking secure quantum key distribution communication links~\cite{Yamamoto1,Yamamoto2,Yamamoto3,Yamamoto4}. In addition, recently, the strong four-wave mixing properties of the frequency tunable limit cycle were used as the {\it fictitious} four-wave mixing medium for the generation of time crystals and optical frequency combs~\cite{HIM23}.

\section{Coexisting Periodic Limit Cycles and their Fractal-Like Basin of Attraction}
\label{sec:fractal}
In this section, we focus on the deterministic behavior of the system in a specific dynamical regime that presents particularly interesting features. This regime corresponds to a finite, constant optical detuning expressed by the parameter $\Delta=\Omega/\omega_r$, where $\omega_r$ is the free-running angular relaxation frequency, which to a good approximation is equal to $\sqrt{2P/T}$~\cite{Herrera2021}.

\begin{figure}[ht!]
\includegraphics[width=\linewidth]{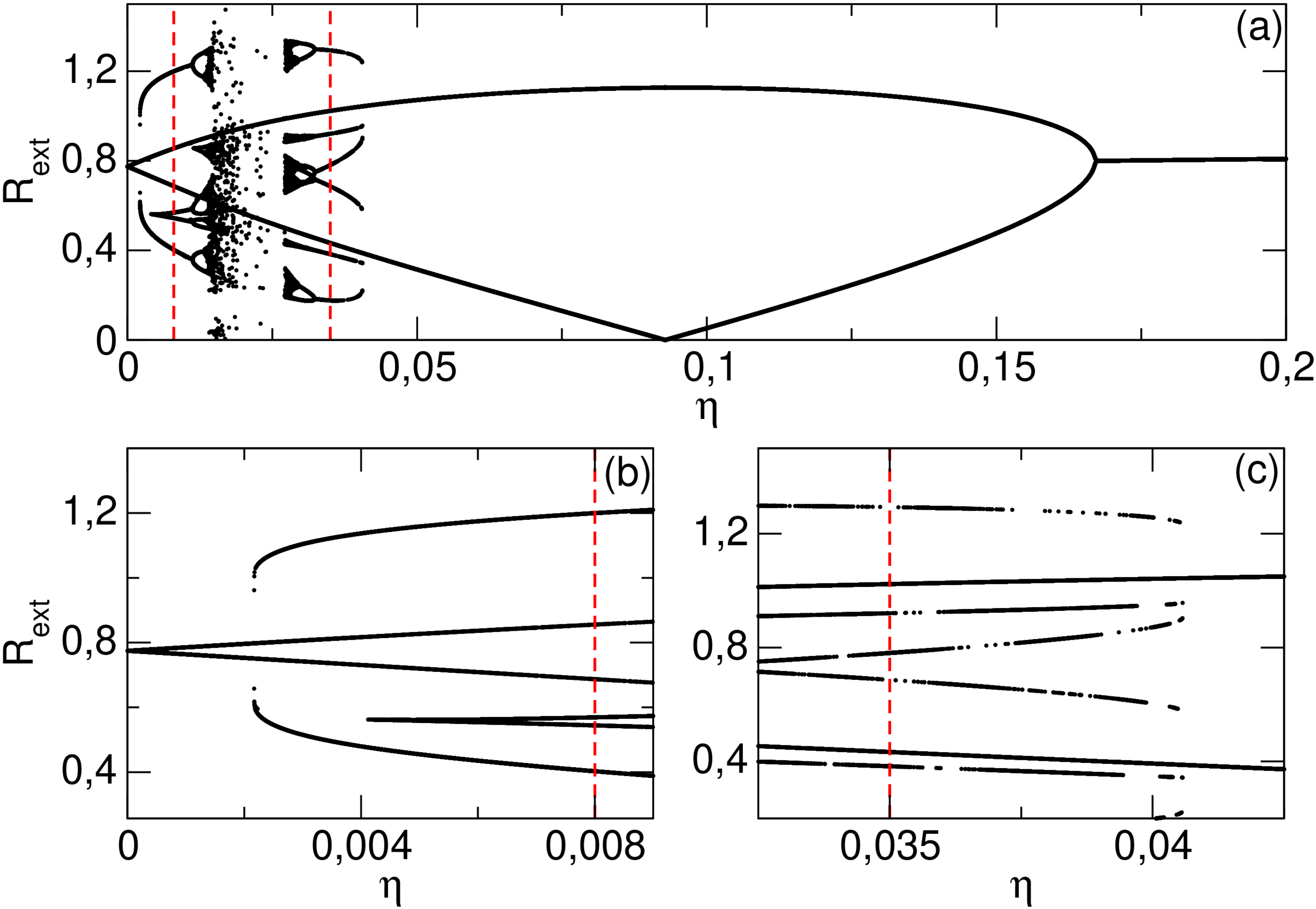}
\caption{(a) Orbit diagram showing the extrema of the electric field amplitude for varying injection strength $\eta$, obtained by numerically integrating the rate equations Eqs.~(\ref{eq:rate_eq_R}-\ref{eq:rate_eq_N}) for the following parameter values: $T=1000$, $\alpha=6$, $P=0.6$, and $\Delta=3$. Panels (b) and (c) show blow-ups around $\eta=0.008$ and $0.035$, respectively, marked by the vertical red dashed lines.}
    \label{fig:orbit_diagram}
\end{figure}

We numerically integrate Eqs.~(\ref{eq:rate_eq_R}-\ref{eq:rate_eq_N}) for $D=0.0$ in time using a standard 
fourth-order Runge-Kutta algorithm. Figure~\ref{fig:orbit_diagram} depicts an orbit diagram that captures the rich complexity of the injection locking architecture for the specific choice of the detuning. In particular, we have plotted the extrema of the amplitude of the electric field in dependence on the injection strength $\eta$. Starting at $\eta=0.2$ where the only solution is a stable fixed point, the system, as $\eta$ decreases, undergoes a Hopf bifurcation at $\eta=0.167$ and a stable limit cycle is born. This is a well-known dynamical scenario of the optically injected laser model that has been extensively studied and experimentally confirmed in multiple types of laser oscillators, including quantum well and quantum dot lasers~\cite{simpson1995} and
recently quantum cascade oscillators~\cite{wang2023}. In a series of papers in the 1990s, we recorded multiple transitions into chaos and coherence collapse states and performed a set of original stochastic and deterministic computations to prove the interplay of stochasticity and nonlinearity~\cite{simpson1995}.
 
\begin{figure}[h!]
    \includegraphics[width=\linewidth]{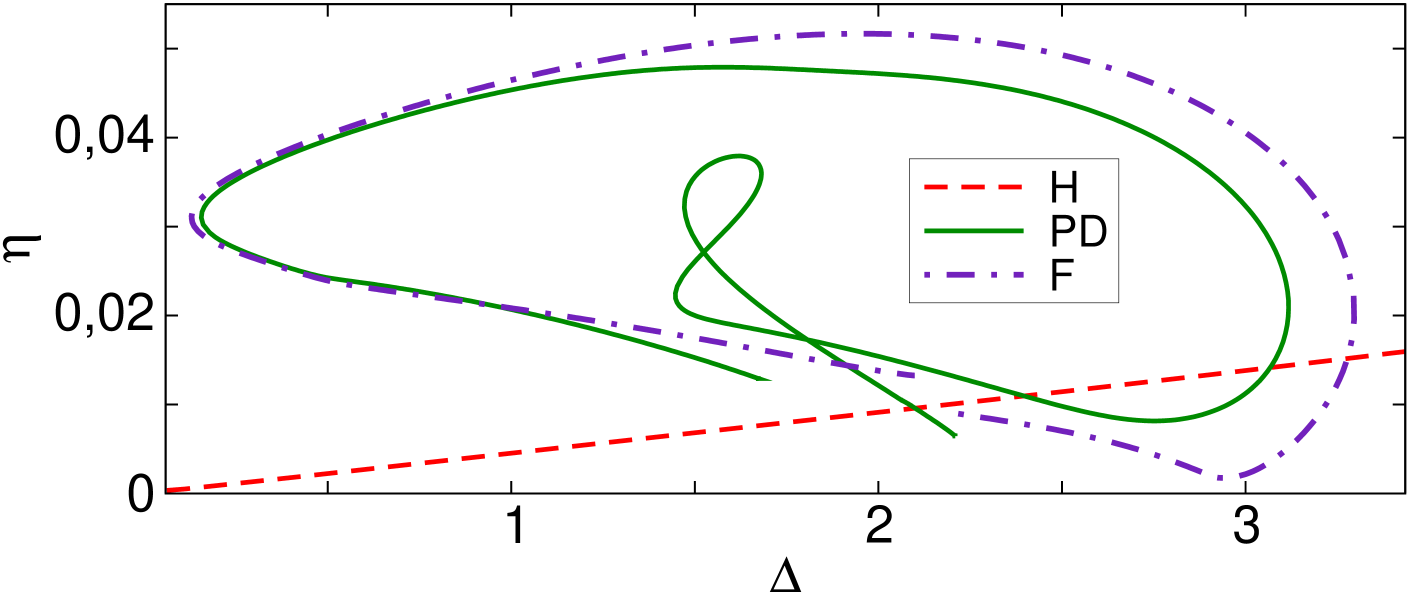}
    \caption{Bifurcation lines obtained by a numerical continuation of Eqs.~(\ref{eq:rate_eq_R}-\ref{eq:rate_eq_N}) in the $(\Delta, \eta)$ parameter plane. ``H" marks the Hopf bifurcation, ``F" the fold (saddle-node) bifurcation of limit cycles, and `PD" the period-doubling bifurcation. Other parameter values are kept fixed to: $T=1000$, $\alpha=6$, $P=0.6$.}
    \label{fig:bif_lines}
\end{figure}

As $\eta$ decreases, a new periodic orbit is born, namely, a period-3 limit cycle as we discuss next, through a saddle-node (fold) bifurcation of limit cycles. 
This occurs at two values of the injection strength, $\eta=0.04$ and $\eta=0.00218$, while for intermediate values of $\eta$ these limit cycles undergo period-doubling bifurcations leading to a narrow chaotic region. A detailed bifurcation diagram showing the aforementioned bifurcation lines in the $(\Delta,\eta)$ parameter space is shown in Fig.~\ref{fig:bif_lines}~\cite{MATCONT}.

Our focus of interest here is two regions for the injection strength ($\eta=0.008$ and $\eta=0.035$), marked by the vertical red lines in Fig.~(\ref{fig:bif_lines}) and blown-up in Figs.~\ref{fig:orbit_diagram}(b) and (c), where two periodic limit cycles coexist.
Specifically, a lower amplitude period-1 (P1) or four-wave mixing (FWM) limit cycle coexists 
with a higher amplitude period-3 (P3) limit cycle, born through a fold bifurcation. 
Notethat the stable P3 limit cycle lives on an isola, that is an isolated branch of solutions which is not
connected to the P1 branch. Isolas have been found on a wide set of optical configurations, including in arrays of nonlinear optical maps~\cite{OHT87}, four-wave mixing optical bistability resonators and three-photon optical excitations of a single cyclotron electron~\cite{KAP85,DIN87}.

\begin{figure}[htbp]
\centering\includegraphics[width=10cm]{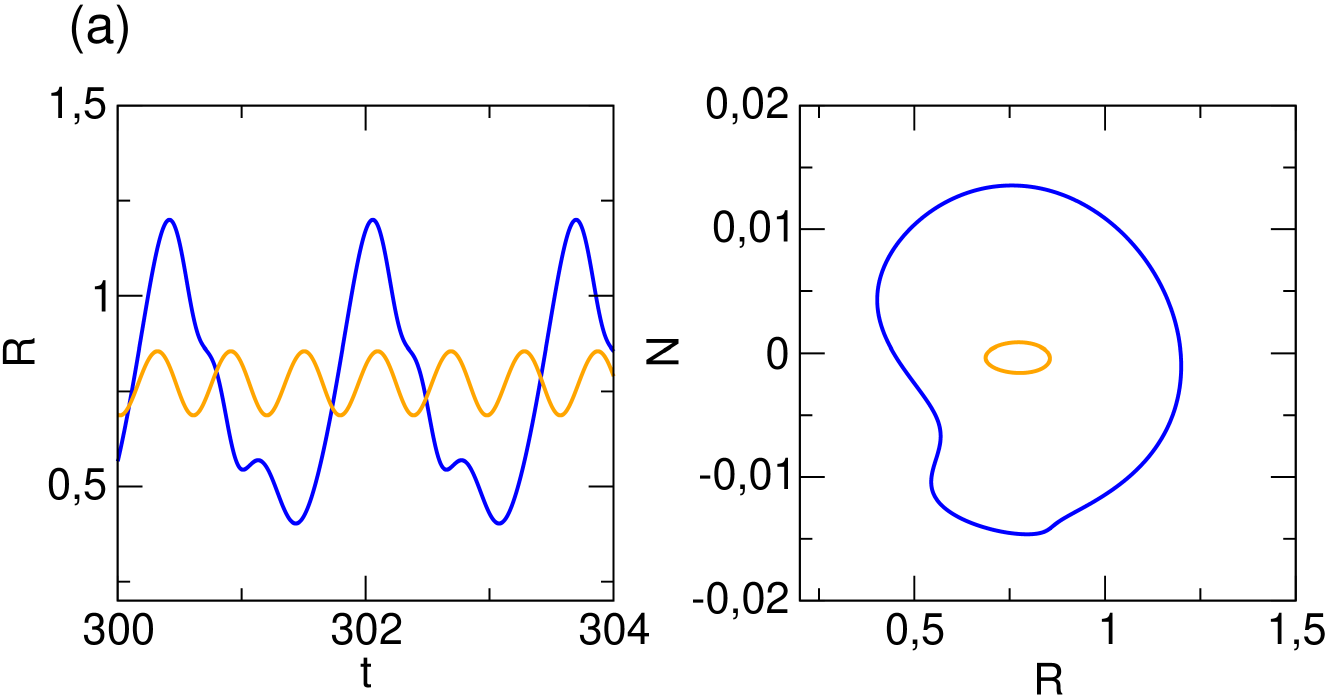}
\centering\includegraphics[width=10cm]{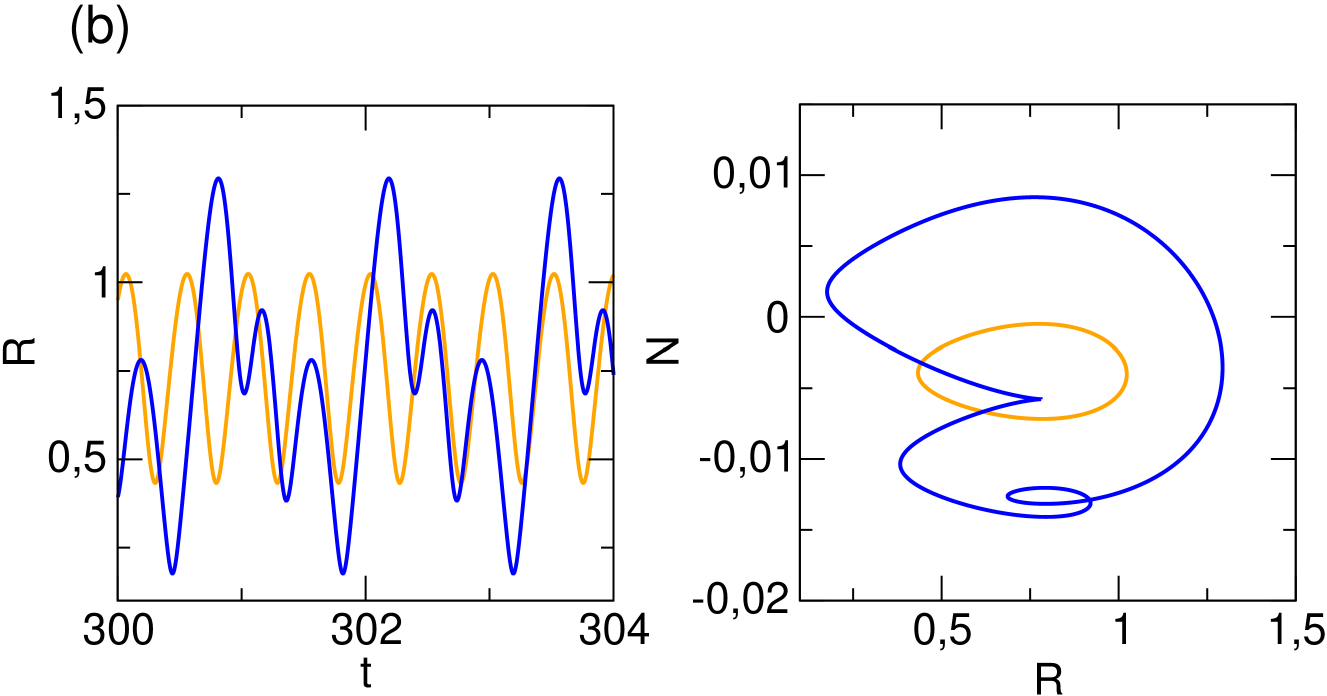}
\caption{Deterministic time series of the electric field amplitude (left) and phase portraits in the $(R,N)$ plane (right) of the two limit cycles coexisting for a value of the injection rate (a) $\eta=0.008$ and (b) $0.035$, marked by the vertical red dashed lines in Fig.~\ref{fig:orbit_diagram}. Blue (dark) color corresponds to the period-3 solution and orange (light) color to the FWM (P1) solution. Other parameter values: $T=1000$, $\alpha=6$, $P=0.6$, $\Delta=3$, and $D=0.0$.}
\label{fig:deterministic_ts_pp}
\end{figure}

The time series of the electric field amplitude (left panel) and phase portraits (right panel) of these orbits are shown in Figs.~(\ref{fig:deterministic_ts_pp})(a) and (b) for $\eta=0.008$ and $\eta=0.035$ in (a) and (b), respectively. For the weak injection case ($\eta=0.008$), notice that compared to the P1 orbit (orange, light), the P3 limit cycle (blue, dark) is about three times larger in oscillation amplitude and occupies a larger region of the phase space $(R,N)$.

\begin{figure}[htbp]
\centering\includegraphics[width=10cm]{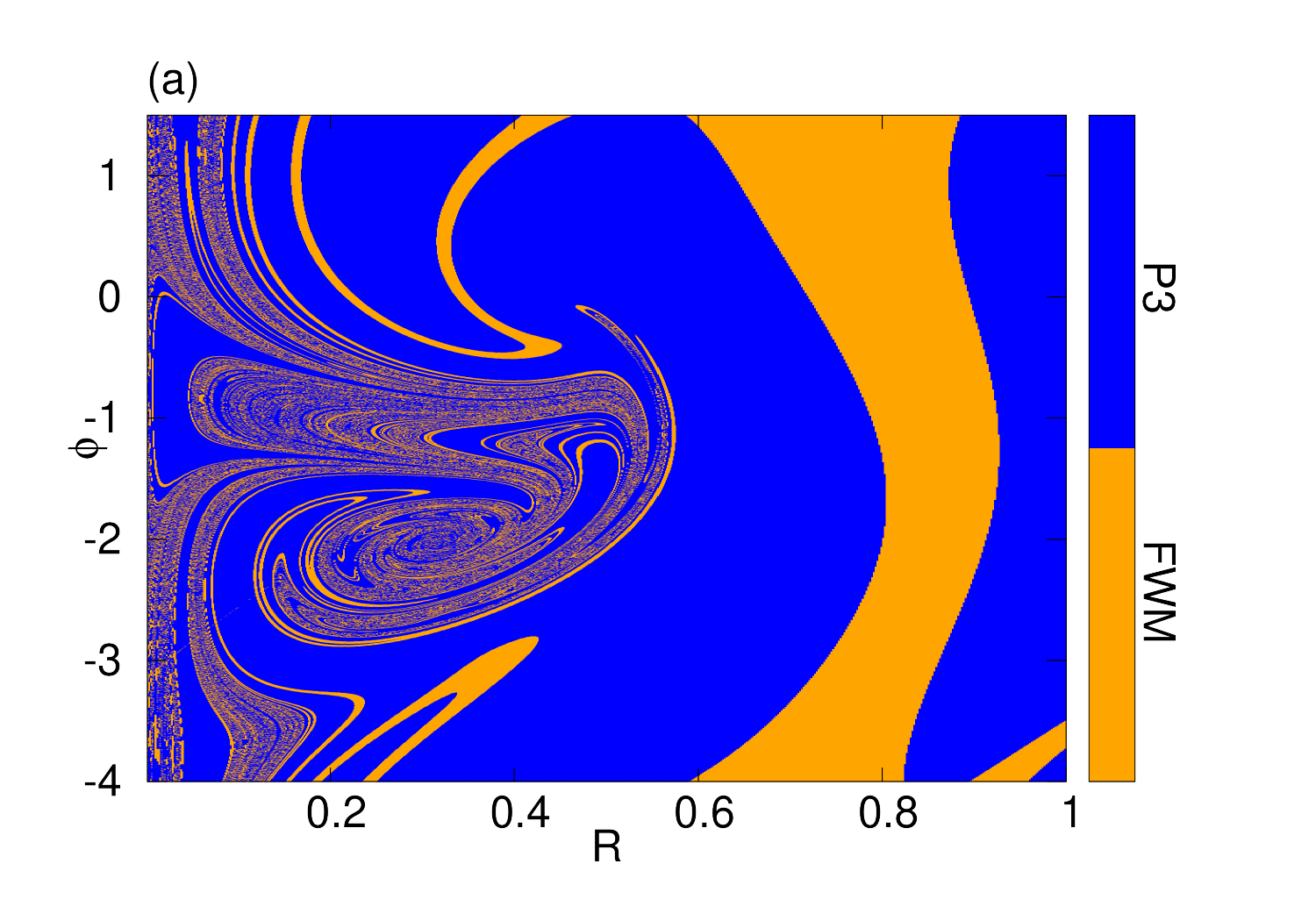}  
\centering\includegraphics[width=10cm]{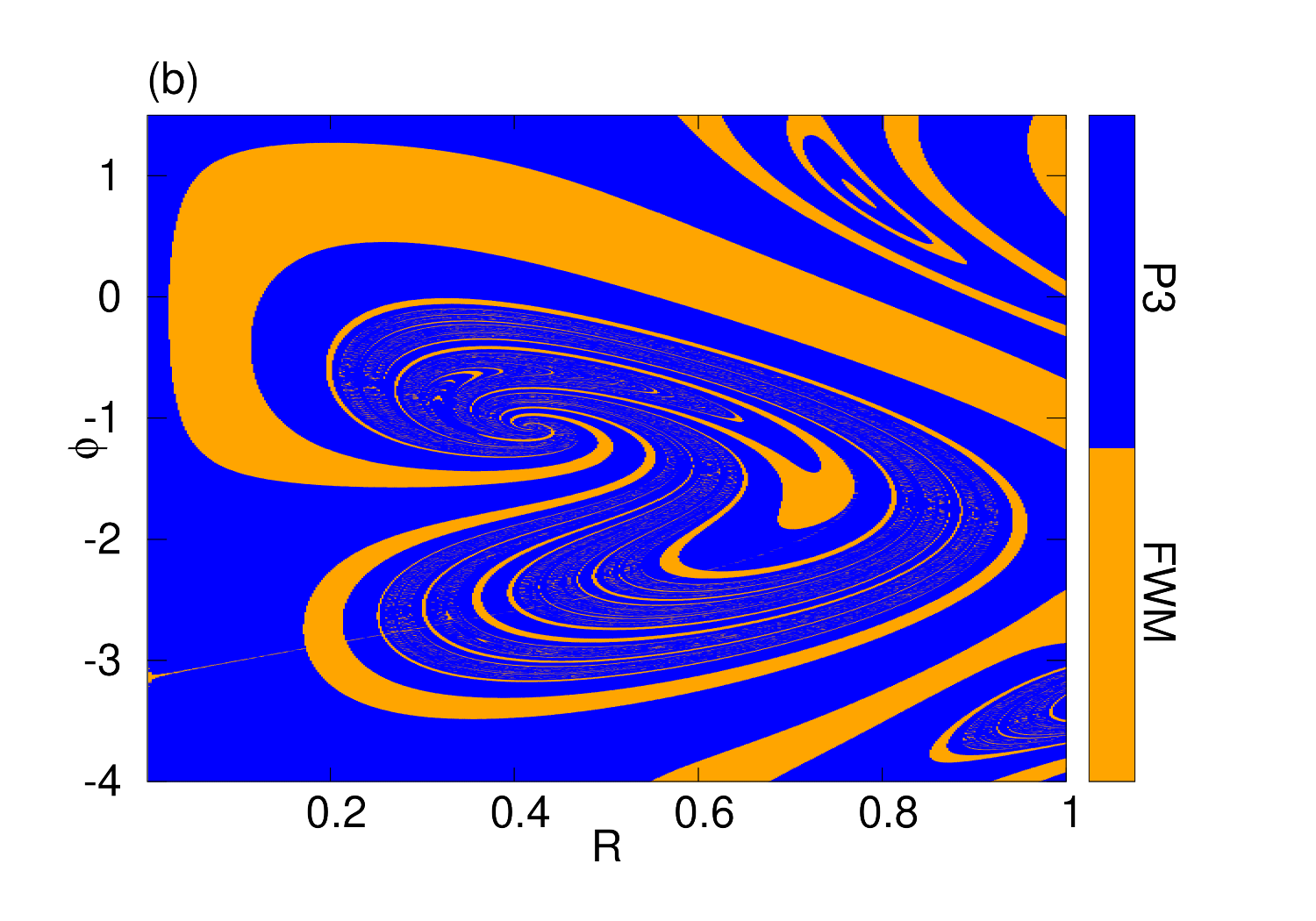}    
\caption{Basins of attraction in the $(R,\phi)$ plane for two values of injection strength (a) $\eta=0.008$ and (b) $0.035$. The initial condition for $N$ is set to zero. Blue (dark) color marks the initial conditions leading to a period-3 solution and orange (light) color marks the initial conditions leading to the coexisting FWM (period-1) solution. Other parameter values are: $T = 1000$, $\alpha = 6$, $P = 0.6$, $\Delta=3$, and $D=0.0$.}.
\label{fig:basins}
\end{figure}

The multistability exhibited by the model naturally renders it very sensitive to the choice of initial conditions. To illustrate this sensitivity, we have identified the basins of attraction of period one and
period-3 oscillatory behavior for both values of the injection strength. 
These are displayed in Figs.~\ref{fig:basins}(a) and (b), respectively, in the $(R,\phi)$ plane. Blue (dark) regions denote the set of initial conditions $(R(0),\phi(0))$ leading to P3 oscillations,
while orange (light) regions are the ones leading to period-1 (FWM) oscillations.  The other state variable is initially fixed to zero, $N(t=0)=0.0$. In both Fig.~\ref{fig:basins}(a) and (b), it is observed that the sets of initial conditions leading to the P3 solution constitute a substantial part of the $(R,\phi)$ plane as indicated by the area occupied by the blue (dark) regions.
Overall, the basins show a fractal-like composition as two long-time behaviors of the limit cycles are mixed and intertwined. This behavior is similar to the P3/chaos region that has been found numerically for zero frequency optical detuning~\cite{Jason2010}, with a key difference: in the latter, the system was locked while in this case, the system is operating in an un-locked FWM region coexisting for large regions of the injection rate with a P3 orbit born out of a saddle mode bifurcation. 

The knowledge of the basins of attraction of coexisting states is essential for determining the usefulness of the system in practical applications. For example, switching between outputs with different intensities and spectral properties may be utilized in optical communications~\cite{LIU20}.
In the following section, we study the effect of noise on the spectral linewidth and lineshape of the two coexisting limit cycles and the respective implications of the numerically found fractality in their basins of attraction.

\section{Optical power spectra and the effect of noise}
\label{ops_noise}
\begin{figure}[]
    \includegraphics[width=10cmh]{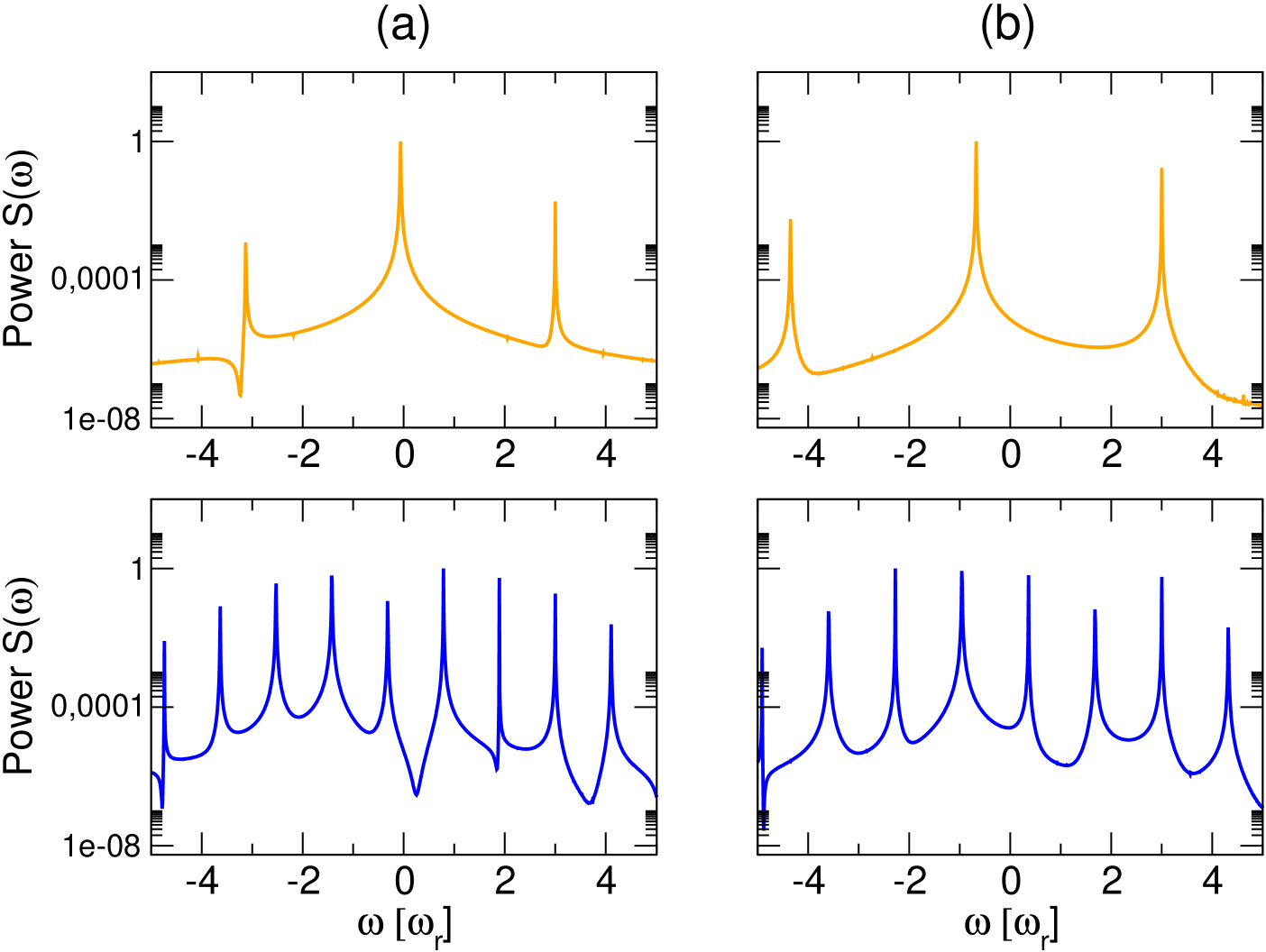}
    \caption{Deterministic optical power spectra for (a) $\eta=0.008$ and (b) $\eta=0.035$. Blue (dark) color corresponds to the optical power spectrum of the period-3 solution (P3) and orange (light) color to the optical power spectrum of the coexisting period-1 (FWM) solution. Other parameter values are: $T=1000$, $\alpha=6$, $P=0.6$, $\Delta=3$, and $D=0.0$.}
    \label{fig:deterministic_spectra}
\end{figure}

The Optical Power Spectrum has been the prime observable to dissect, understand, and design the long-time behavior of optically coupled lasers. This fundamental quantity combines the amplitude and phase fluctuations of the recorded radiation emitted out of the slave laser cavity, where the noise of the laser amplifying medium as well as the nonlinear optical gain are encoded in the output radiation. Its numerical calculation is done based on the Fourier
transform of the electric field $E(t)$, given by:
\begin{equation}
    S(\omega)=\left |\int_{-\infty}^{+\infty} R(t)e^{i\phi(t)}e^{i\Omega t}e^{-i\omega t} dt \right |^2.
\end{equation}
The deterministic power spectra (normalized to their maximum value) of the coexisting FWM (top panel, orange) and P3 (bottom panel, blue) orbits are depicted in Fig.~\ref{fig:deterministic_spectra} for both values of the injection strength, (a) $\eta=0.008$ and (b) $\eta=0.035$.
As we anticipate, in the case of the period-1 limit cycle, apart from the power concentration around $\omega=0$, there is a prominent peak at the injected frequency $\omega=\Delta \omega_r$ ($\Delta=3$). On the other hand, in the case of the P3 limit cycle, we observe strong
power concentrations at integer multiples of the free-running relaxation frequency $\omega_r$.
Moreover, the linewidth of its center line is sharper than that of the corresponding FWM spectrum.

The basins of attraction serve as a guideline for obtaining the FWM or P3 spectrum on demand. However, in real-world conditions, noise is always present.
The key question here is how noise affects the linewidth and the lineshape of two coexisting periodic limit cycles and what is the role of the numerically found fractality in their basins of attraction.
We typically estimate the value of the noise intensity $D$ by connecting it with the linewidth of the center line of the FWM optical power spectrum and keep this value fixed to $0.001$ across all stochastic numerical simulations.

The stochastic differential equations are integrated numerically by applying Milstein's method \cite{TOR14} which is suitable for systems that are driven by both additive and multiplicative noise terms in this case in the amplitude and phase variable of Eqs.~(\ref{eq:rate_eq_R}) and~(\ref{eq:rate_eq_phi}), respectively. The addition of noise helps the system alternate between period-1 and period-3 oscillatory motion, depending on the initial condition imposed by the random seed of each realization.

\begin{figure}[h!]
    \includegraphics[width=10cm]{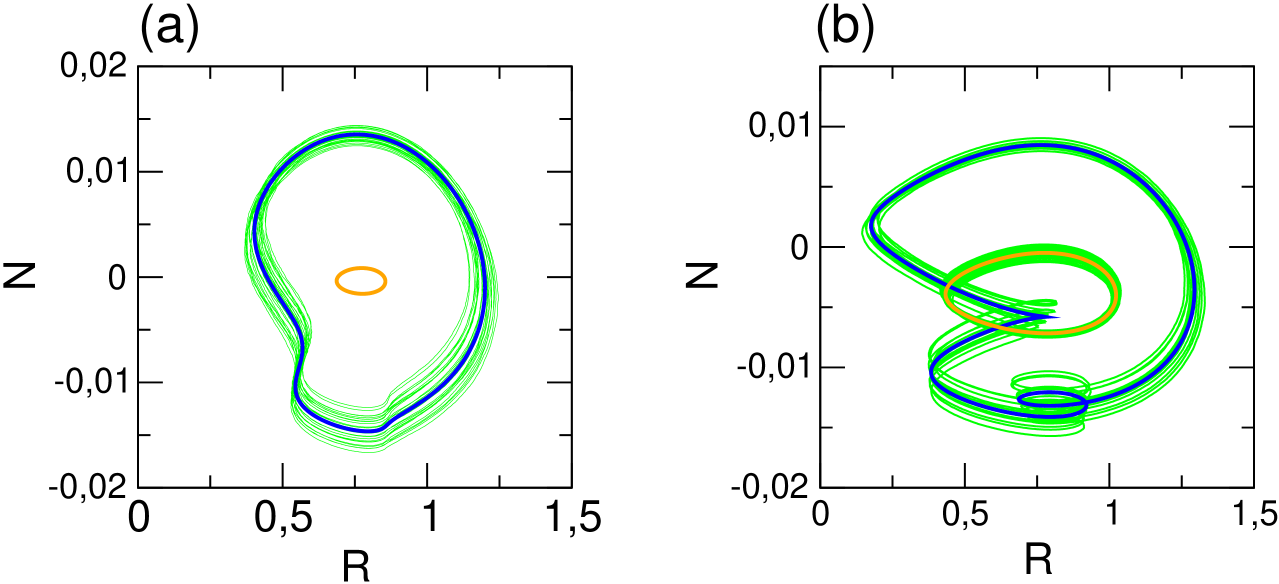}
    \caption{Noisy (green) phase portraits in the $(R,N)$ plane for injection rate (a) $\eta=0.008$ and (b) $\eta=0.035$.
    The corresponding deterministic orbits are plotted on top: Blue (dark) color marks the P3 solution and the orange (light) color marks the FWM solution. Other parameter values are: $T = 1000$, $\alpha = 6$, $P = 0.6$, $\Delta=3$, and $D=0.001$.}  
    \label{fig:noisy_pp}
\end{figure}

The corresponding noisy phase portraits are shown in Fig.~\ref{fig:noisy_pp} in green color, while the deterministic FWM and P3 orbits are also plotted on top for comparison.

\begin{figure}[h!]
    \includegraphics[width=10cm]{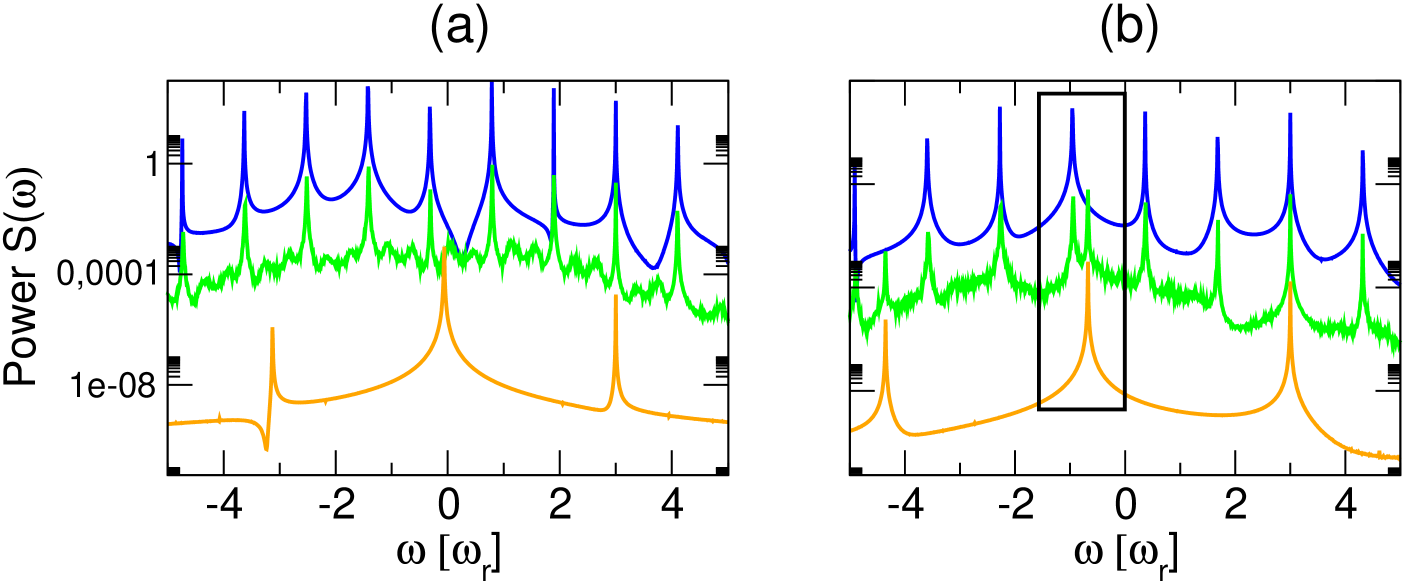}
    \caption{Noisy optical power spectrum for injection rate (a) $\eta=0.008$ and (b) $\eta=0.035$.
     The deterministic optical power spectra are included for comparison (shifted, for visualization purposes): Blue (dark) color marks the optical power spectrum of the P3 solution and the orange (light) color marks the optical power spectrum of the FWM solution.  Other parameter values are: $T = 1000$, $\alpha = 6$, $P = 0.6$, $\Delta=3$, and $D=0.001$.}  
    \label{fig:noisy_spectra}
\end{figure}

Next, we calculate the average spectrum over 100 noise realizations (each with a different random initial condition) comprising 1000 periods in total. These are plotted in Fig.~(\ref{fig:noisy_spectra}), where the deterministic spectra are superimposed 
(and vertically shifted for visualization purposes), for the sake of comparison.

For $\eta=0.008$ (Fig.~\ref{fig:noisy_spectra} (a)), the noisy spectrum is similar to the noise-free P3 (blue, dark) spectrum since, as we discussed above, for weak injection strength noise favors this solution. Naturally, the primary spectrum peaks are thicker, and in addition,
secondary smaller ones appear in between, due to the  
small fluctuations of the noise source. This was partly captured by the experimental measurement revisited in the Appendix A,
however, in that work, the optical power spectrum of the FWM signal was also recorded.
The latter is predicted by our simulations too, but only as a very long transient phenomenon.
On the other hand for $\eta=0.035$, the noisy spectrum reveals the primary peaks of both the FWM (orange, light) and the P3 (blue, dark) solution, marked by the black rectangle in Fig.~\ref{fig:noisy_spectra} (b). This ``mixed" spectrum image reflects the coexistence of the two orbits which for higher injection strengths persists, despite the fluctuations. This is a very interesting feature that was not observed in the reference experiment described in the Appendix A which is definitely worth further investigation in a laboratory setting.

\section{Conclusions and Future Work}
\label{sec:conclusions}
In this manuscript, we have systematically dissected the P1 and P3 limit cycles in the vicinity of optical detuning $\Delta=3$ for a wide region of the injection rate of the master laser. 
Deep numerical scanning of the basin of attraction of these two limit cycles reveals that the phase space has fractal features, making the system unpredictable and posing novel questions about the shape and the thickness of the linewidth of such emitted radiation where, in the presence of noise, the signal is a combination and/or fusion of the two outcomes. We intend to further investigate such types of questions on the interplay of fractality and noise on the long-time optical spectra as well as how to engineer the basins and the strength of the noise to shape the spectral emission for a useful set of applications, such as coherent LIDAR, optical metrology and spectroscopy, and interferometric optical sensing. It is to be noted that key issues of current applications of quantum sensing are deeply dependent on the use of properly sharp RF linewidth as well as wide frequency low noise tunable photonic oscillators.
A key finding on this manuscript is the appealing observation that in the region of parameter space that we examine ($\Delta=3$ and $\eta=0.008$) is that the P3 limit cycle draws most of the trajectories on its path. The P3 is born out of a saddle node bifurcation and with about three times more intensity than the P1 cycle.
Note that very similar behavior can be observed in other scenarios of coexisting limit cycles as well, as in the case of low detuning analysed in Appendix B.
Finally we use these numerical observations to connect and contrast the injected locked photonic
oscillators literature with observations and analysis of supernarrow spectral peaks on the spectrum
on nonlinear microwave oscillators and driven nonlinear micromechanical oscillators.
In the future we will focus on isolas existing on two critical regions of the FWM territory, the
strongly hysteretic vicinity of the subharmonic resonance ($\Delta=2$) and the region of the primary
resonance ($\Delta= 1$), where we also expect strong interplay of the noise and the isolas on the
shape of the optical power spectrum, even substantial sharpening of the optical and RF linewidth
of the spectra.

\subsection*{Acknowledgements}
The VK research portfolio is supported via a VT Innovation Campus start-up,  VT National Security Start-up, and generous gifts from private corporations including IDQ from Geneva to the Virginia Tech Foundation. Funding was provided by NSF Award Abstract \#2328991
under ExpandQISE: Track 1: Prairie View A\&M University-Virginia Tech Partnership in Quantum Science \& Engineering Research and Education.

\subsection*{Disclosures}
The authors declare no conflicts of interest.
 % Disclosures should be listed in a separate nonnumbered section at the end of the manuscript. List the Disclosures codes identified on the \href{https://opg.optica.org/submit/review/conflicts-interest-policy.cfm}{Conflict of Interest policy page}, as shown in the examples below:

\section*{Appendix A}
\label{sec:appendix_A}
The work that motivated our study appeared in a book of abstracts of the annual Optical Society of America Conference in 1996 in Anaheim California~\cite{Varangis1996},
as a rudimentary report on the combined theoretical and experimental study of the coexisting FWM and P3 limit cycles in the dynamical regime of optical detuning $\Delta=3$. Here, we briefly revisit the experimental setup and the main findings which are compared to the results of the present manuscript, in detail, in the main text.

\begin{figure}[h!]
   % \centering
    \includegraphics[width=\linewidth]{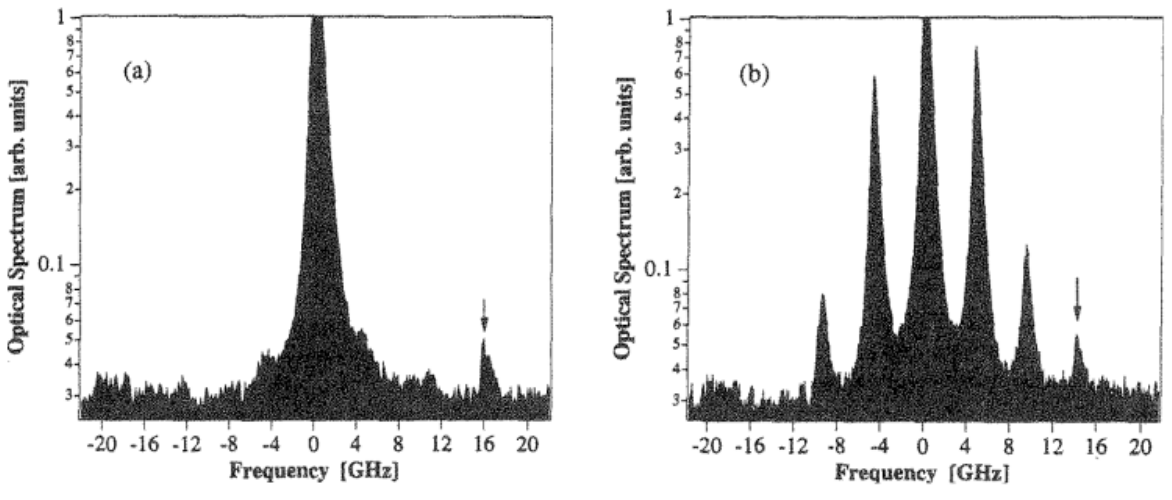}
    \caption{Experimentally recorded optical power spectra of an optically injected quantum well laser of HLP400 type, They were recorded by the scanning of the Fabry-Perot interferometer. The zero frequency in this figure is taken at the shifted center line.  The master injection laser line is on the right marked by an arrow (positive detuning): (a) the FWM signal and (b) the P3 signal. Taken from~\cite{Varangis1996}}.
    \label{fig:exp_opt_spec}
\end{figure} 

The experimental setup used in \cite{Varangis1996} consisted of two single longitudinal mode quantum well lasers, serving as master and slave, of HLP1400  type,  emitting at a wavelength of 830 nm. The laser bars were temperature and current-stabilized and an optical isolator was used to prevent mutual 
light injection between the master and the slave elements. The radiation output of the lasers was monitored with the use of a Newport SR-240 scanning Fabry-Perot interferometer having a free spectral range (FSR) of 8 THz and linewidth resolution of 0.8 GHz. In Fig.~\ref{fig:exp_opt_spec} (respectively Fig. 2 in \cite{Varangis1996}) we observe a typical four-wave mixing signal consisting of three emission lines: the slave center line, the four-wave mixing signal, and the regenerative line at the injected frequency. In Fig.~\ref{fig:exp_opt_spec} (b) we see in addition the appearance of a set of strong side bands at approximately 1/3 $f_{inj}$ and 2/3 $f_{inj}$. These sidebands are the manifestation of the period three isolated period solution, coexisting with the P1 signal. 

\section*{Appendix B}
\label{sec:appendix_B}
To a large extent, the findings reported in this work are not specific to the aforementioned fixed parameter values but, rather, extend to other regions of the parameter space as well. In the following we demonstrate the case of very low detuning (close to locking $\Delta=0.1$) and different linewidth enhancement factor and pump current ($\alpha=4, P=1.0$). For these parameter values the system undergoes a different dynamical scenario, however, there is clear evidence of coexisting periodic attractors, which in this case are both period-1 limit cycles, called LC1 and LC2 hereafter.
\begin{figure}[h!]
    \includegraphics[width=\linewidth]{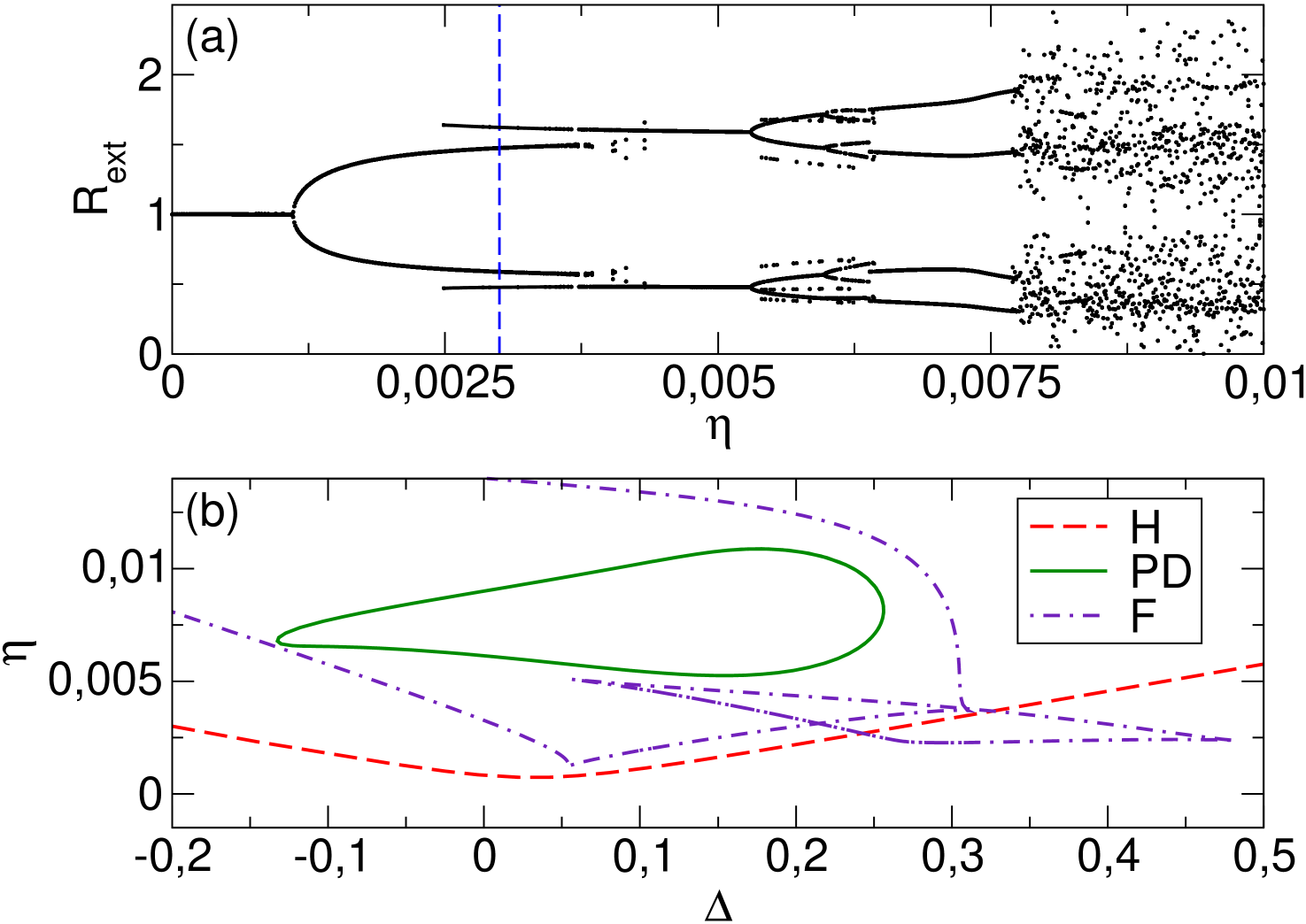}
    \caption{(a) Orbit diagram showing the extrema of the electric field amplitude for varying injection strength $\eta$, obtained by numerically integrating the rate equations Eqs.~(\ref{eq:rate_eq_R}-\ref{eq:rate_eq_N}) for $\Delta=0.1$. The vertical blue dashed line marks the injection strength value where the periodic attractors LC1 and LC2 coexist. (b) Bifurcation lines obtained by a numerical continuation of Eqs.~(\ref{eq:rate_eq_R}-\ref{eq:rate_eq_N}) in the $(\Delta, \eta)$ parameter plane. ``H" marks the Hopf bifurcation, ``F" the fold (saddle-node) bifurcation of limit cycles, and `PD" the period-doubling bifurcation. Other parameter values are kept fixed to: $T=1000$, $\alpha=4$, $P=1.0$.}
    \label{fig:appB_fig1}
\end{figure}

This is shown in the orbit diagram of Fig.~\ref{fig:appB_fig1}(a) where we have plotted the extrema of the amplitude of the electric field in dependence on the injection strength $\eta$. Starting at $\eta=0.0$ where the only solution is a stable fixed point, the system, as $\eta$ increases, undergoes a Hopf bifurcation at $\eta=0.0012$ and a stable limit cycle (LC2) is born. For $\eta>0.025$ there is a coexisting branch of solutions (LC1) that has been born through a fold bifurcation and lives an isola like we have already seen in the main paradigm. A detailed bifurcation diagram in the $(\Delta,\eta)$ plane is shown in Fig.~\ref{fig:appB_fig1}(b) where the complex structure of consecutive fold and period-doubling instabilities are shown as the injection strength increases. 
Here we will focus on region where LC1 and LC2 coexist and therefore fix the injection strength to a constant value $\eta=0.003$, marked by the blue dashed line in Fig.~\ref{fig:appB_fig1}(a).

\begin{figure}[]
\centering\includegraphics[width=\linewidth]{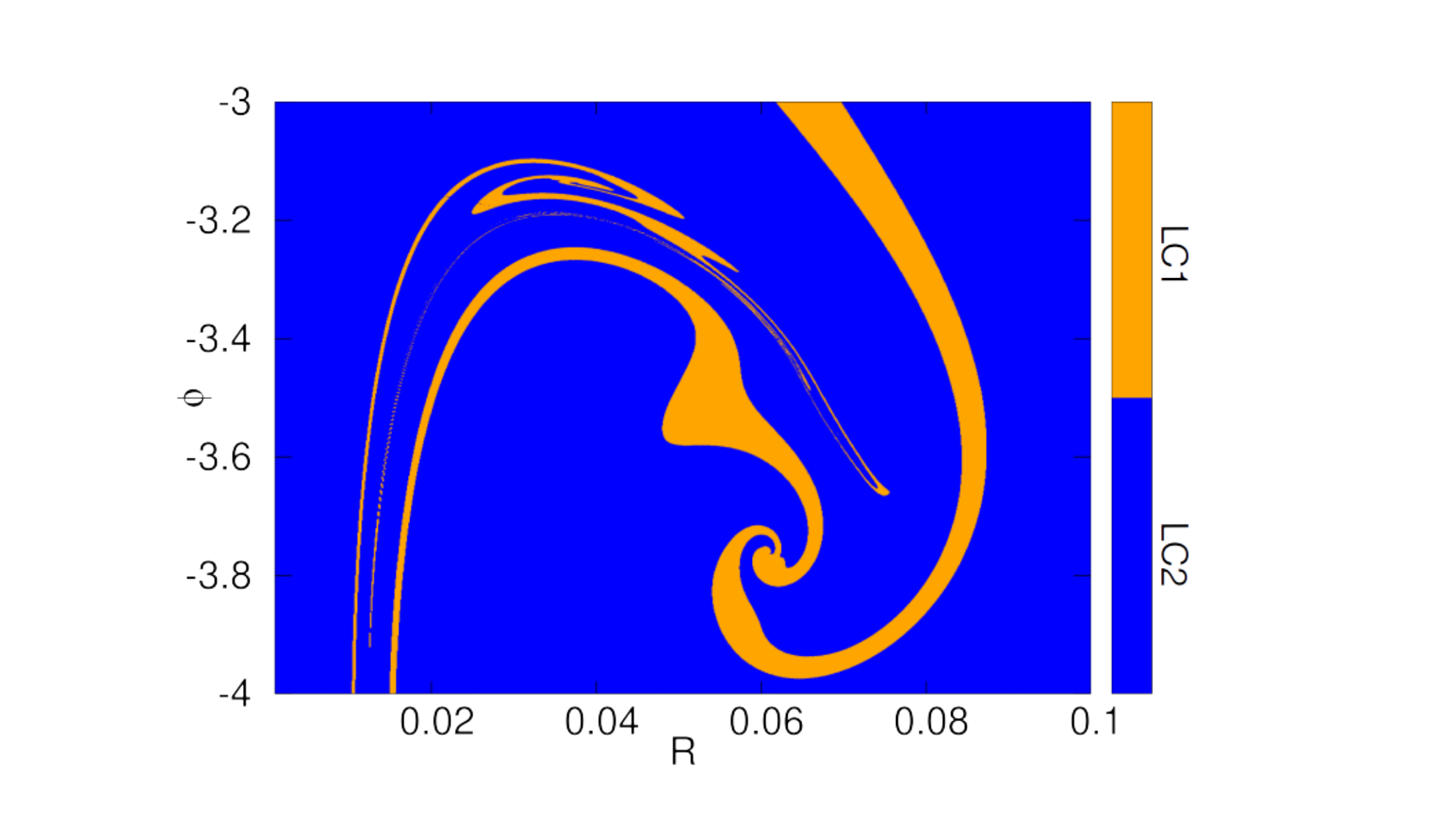}  
\caption{Basins of attraction in the $(R,\phi)$ plane for $\eta=0.003$ (marked by the blue dashed line in Fig.~\ref{fig:appB_fig1}(a)). The initial condition for $N$ is set to zero. Blue (dark) color marks the initial conditions leading to the LC2 solution and orange (light) color marks the initial conditions leading to the LC1 solution. Other parameter values are: $T=1000$, $\alpha=4$, $P=1.0$, $\Delta=0.1$, and $D=0.0$.}.
\label{fig:appb_fig2}
\end{figure}

The next step is to identify the basins of attraction of LC1 and
LC2 oscillatory behavior. 
These are displayed in Fig.~\ref{fig:appb_fig2}, in the $(R,\phi)$ plane. Blue (dark) regions denote the set of initial conditions $(R(0),\phi(0))$ leading to LC2 oscillations,
while orange (light) regions are the ones leading to LC1 oscillations.  The other state variable is initially fixed to zero, $N(t=0)=0.0$. We observed that the sets of initial conditions leading to the LC2 solution constitute a substantial part of the $(R,\phi)$ plane as indicated by the area occupied by the blue (dark) regions.
Similarly to Sec.~\ref{sec:fractal}, the basins show a fractal-like composition as two long-time behaviors of the limit cycles are mixed and intertwined. 

\begin{figure}[]
    \includegraphics[width=\linewidth]{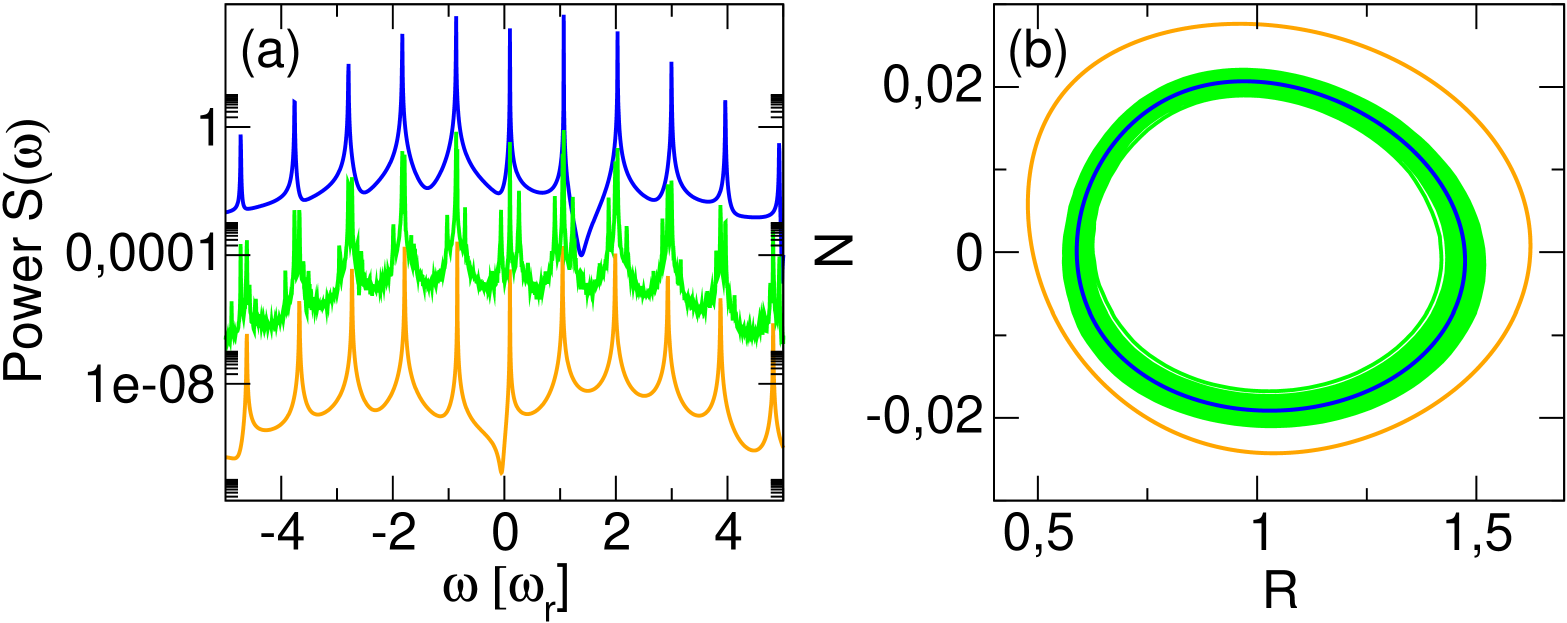}
    \caption{(a) Noisy optical power spectrum and (b) corresponding phase portrait (green) in the $(R,N)$ plane for injection rate $\eta=0.003$. The deterministic optical power spectra are included in (a) for comparison (shifted, for visualization purposes): Blue (dark) color marks the optical power spectrum of the LC2 solution and the orange (light) color marks the optical power spectrum of the LC1 solution. Similarly, in (b) the deterministic orbits are plotted on top: Blue (dark) color marks the LC2 solution and the orange (light) color marks the LC1 solution. Other parameter values are: $T = 1000$, $\alpha = 4$, $P = 1.0$, $\Delta=0.1$, and $D=0.001$.}  
    \label{fig:appB_fig3}
\end{figure}

We expect this to be reflected in the stochastic response of this particular dynamical regime. Indeed, for weak noise the LC2 solution is favored, which is illustrated in Fig.~\ref{fig:appB_fig3}(b) where the noisy phase portrait is shown in green color, while the deterministic LC1 (orange) and LC2 orbit (blue) are also plotted on top for comparison. 
Finally, with respect to the spectral properties the situation here is quite different, primarily owing to the fact that the LC1 and LC2 attractors have almost identical frequencies, differing only in amplitude, and therefore their optical power spectra are effectively indiscernible.   
We have calculated the average spectrum over 100 noise realizations (each with a different random initial condition) comprising 1000 periods in total and plotted it in Fig.~\ref{fig:appB_fig3}(a), where the deterministic spectra are superimposed 
(and vertically shifted for visualization purposes). 
As expected, the primary spectrum peaks are thicker, and in addition,
secondary smaller ones appear around and in between them, due to the  
small fluctuations of the noise source. 

\end{document}